# Low-Drift Flow Sensor with Zero-Offset Thermopile-Based Power Feedback

M. Dijkstra, T.S.J. Lammerink, M.J. de Boer, J.W. Berenschot, R.J. Wiegerink and M. Elwenspoek
MESA+ Institute for Nanotechnology, University of Twente
PO Box 217, 7500 AE Enschede, The Netherlands
Tel: +31-53489-4438, Fax: +31-53489-3343, E-mail: m.a.dijkstra@ewi.utwente.nl

*Abstract* – A thermal flow sensor has been realised consisting of freely-suspended silicon-rich silicon-nitride microchannels with an integrated Al/poly-Si$^{++}$ thermopile in combination with up- and downstream Al heater resistors. The inherently zero offset of the thermopile is exploited in a feedback loop controlling the dissipated power in the heater resistors, eliminating inevitable influences of resistance drift and mismatch of the thin-film metal resistors. The control system cancels the flow-induced temperature difference across the thermopile by controlling a power difference between both heater resistors, thereby giving a measure for the flow rate. The flow sensor was characterised for power difference versus water flow rates up to 1.5 μl·min$^{-1}$, being in good agreement with a thermal model of the sensor, and the correct low-drift operation of the temperature-balancing control system has been verified.

## I. INTRODUCTION

The miniaturisation of microfluidic components in chemical analysis, synthesis, biotech and nanotechnology asks for the accurate and reliable measurement of tiny fluid flow rates in the order of nl·min$^{-1}$. Current micromechanical thermal flow sensors are capable of measuring liquid flows down to a few nl·min$^{-1}$ [1, 2]. However, the long-term measurement accuracy of these sensors is ultimately limited by the notorious problem of drift in the properties of temperature sensing and heating elements, for which thin-film metal layers are commonly used.

This paper presents a thermal flow sensor and control system that exploits the differential temperature-dependent output $V_{TC}$ of a thermopile in order to obtain a flow sensor that is independent of ambient temperature, resistance drift and mismatch of thin-film metal resistors used (Fig. 1). The dissipated powers $P_1$, $P_2$ in the heater resistors up- and downstream from the thermopile are determined and controlled independently of resistance values by measuring both the current through and the voltage across the resistors, thereby eliminating influences of resistance drift. Additionally, the control-system feedback loop is made independent on non-linear thermopile characteristics $m_{TC}(\Delta T_{TC})$ and drift in thermocouple material properties by cancelling the generated output voltage $V_{TC}$, providing for an inherently drift-free zero-offset error signal (Fig. 1).

The output voltage $V_{TC}$ and temperature difference across the thermopile is cancelled by controlling a power difference $\Delta P$ between the two heater resistors, with a fixed amount of total power $P_T$ being dissipated, according to the temperature-balancing anemometry principle [3, 4]. Applying a fluid flow changes the temperature distribution around the heaters by forced convection. This results in the relation $\Delta P/P_T$ being dependent on the flow rate $Q$. A measure of the flow rate $m_{TB}(\Delta P/P_T)$ can be obtained, with $\Delta P/P_T$ being linear for small flow rates.

## II. SENSOR FABRICATION

The fabricated thermopile flow sensor incorporates mechanically strong freely-suspended surface microchannels for on-chip transport of fluid. Surface microchannels are fabricated without the requirement for a sacrificial layer and allow for the integration of sensor elements in close proximity to the fluid [5].

Fig. 2 gives a schematic overview of the process scheme for the fabrication of the thermopile flow sensor. Surface microchannels are created by isotropic dry etching, using high-density $SF_6$ plasma with zero self-bias, through etch holes 2 μm in width, in a low-stress 500 nm silicon-rich silicon-nitride (SiRN) layer. The etch holes and inner surfaces of the microchannels are conformally coated by a second low-stress LPCVD deposited 1.2 μm SiRN layer, resulting in completely sealed microchannels, while leaving a planar substrate surface for the integration of an Al/poly-Si$^{++}$ thermopile and Al heater resistors. The thermopile is created by LPCVD deposition and boron doping by solid source diffusion of a 200 nm poly-Si$^{++}$ layer and sputtering of a 200 nm Al layer (Fig. 2a).

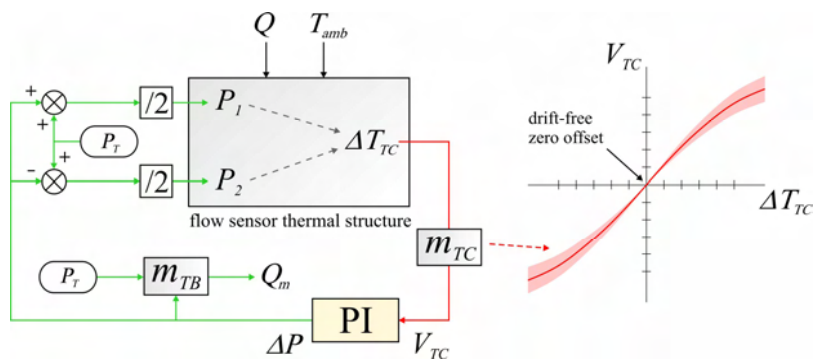

Fig. 1. Zero-offset thermopile-based temperature-balancing flow sensor.

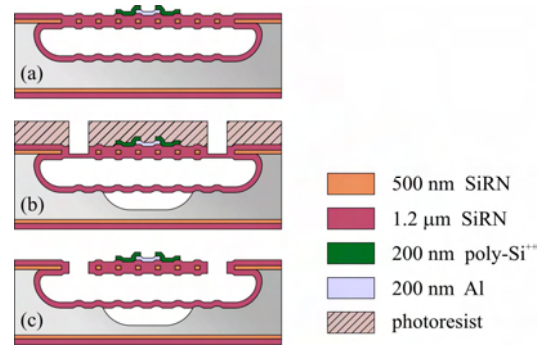

Fig. 2. Process scheme for the fabrication of the microchannel thermopile flow sensor.

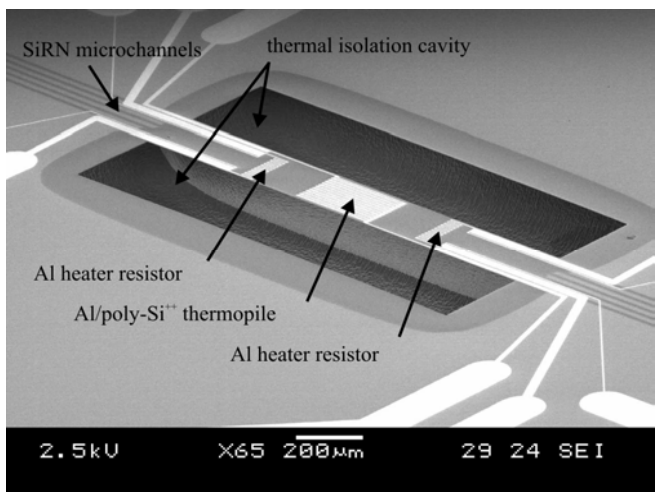
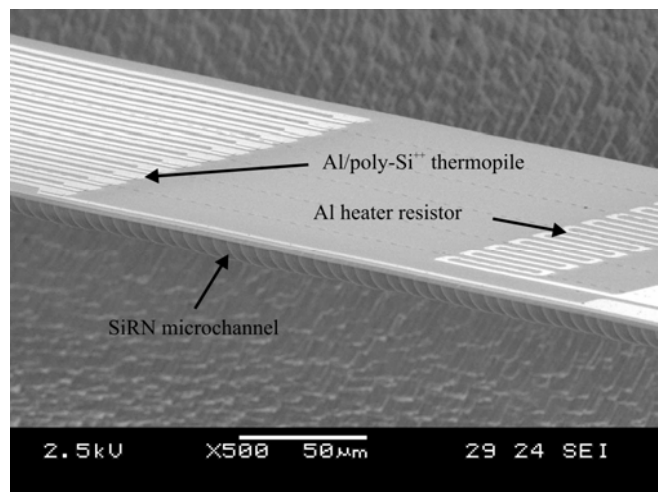

Fig. 3. Microchannel flow sensor with integrated Al heater resistors and Al/poly-Si$^{++}$ thermopile.

The surface microchannels are released by $SF_6$ plasma etching, for thermal isolation from the heat-conducting substrate, with the photoresist mask protecting the sensor elements during the release (Fig. 2b). The photoresist is removed after fluidic entrance holes are etched through the SiRN layer, allowing for the direct interfacing to the microchannels on the substrate surface.

Fig. 3 shows SEM micrographs of the fabricated thermopile flow sensor. The sensor contains five parallel 20 μm diameter microchannels, spanning a 1.6 mm long thermal isolation cavity. The parallel microchannels have 200 μm total width, which allows for the integration of an Al/poly-Si$^{++}$ thermopile with 26 junctions and up- and downstream Al heater resistors. The Al heater resistors have four-point measurement contacts for accurate resistance measurements.

### III. SENSOR MODELLING

A two-dimensional cylindrical finite volume method (FVM) thermal model of the sensor was constructed in MATLAB using sparse-matrix calculations. The model gives approximate solutions to the temperature field surrounding the freely-suspended microchannel, taking into account the thermal conductivities of the freely-suspended microchannel containing fluid, the sensor elements, electrical connection leads and the surrounding air. Ambient temperature is imposed as a boundary condition at a radius of 450 μm from the microchannel, modelling the under-etched thermal-isolation cavity. The temperature difference between both sides of the thermopile is made zero after successive model iterations finding the power difference $\Delta P$ required for a particular flow rate.

Fig. 4 shows the temperature field at 0.3 μl·min$^{-1}$ water flow, with 0.1 mW total power $P_T$ being dissipated. Several temperature profiles, along the freely-suspended microchannel, were obtained from model calculations at different flow rates (Fig. 5). The temperature profile at 0 μl·min$^{-1}$ is symmetric, with both heaters dissipating an equal amount of power. The temperature profile around the heaters is disturbed by forced convection on applying water flow through the microchannel. Maintaining zero temperature difference between both sides of the thermopile $\Delta T_{TC}$ requires the upstream heater to dissipate less power than the downstream heater.

The upstream heater temperature decreases with increasing flow rate and decreasing heating power. However, the downstream heater does not increase in temperature for water flow rates higher than 0.6 μl·min$^{-1}$, due to the total power $P_T$ being limited, while a large amount of heat is advected from the heater, limiting the sensitivity of the flow sensor for flow rates higher than 0.6 μl·min$^{-1}$.

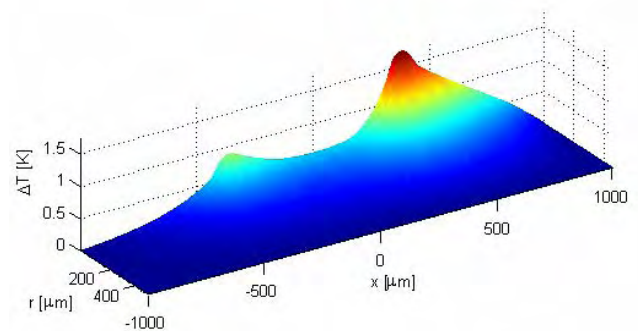

Fig. 4. FVM thermal model result showing the temperature field at a flow rate of 0.3 μl·min$^{-1}$.

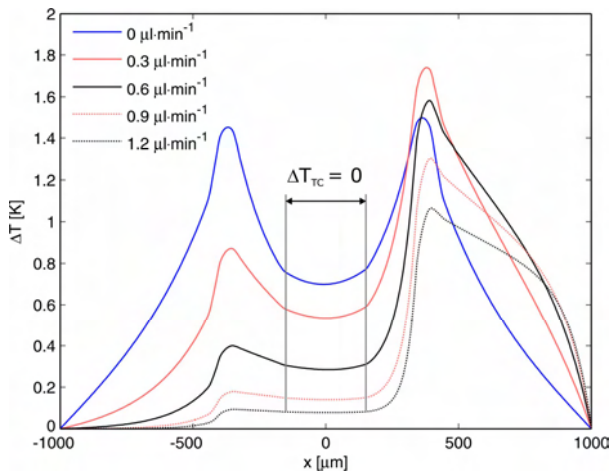

Fig. 5. FVM model calculation results giving temperature profiles along the microchannel.

## IV. EXPERIMENTAL RESULTS

Thermopile flow sensor chips were fabricated $12.5 \times 12.5$ mm in size, with fixed bondpad and microchannel entrance hole positions for self alignment with a chip holder, containing O-rings for fluidic interfacing and twelve pogo-pins for electrical connections to the sensor chip (Fig. 6).

An elevation head up to a few meters $\Delta h$ was used to force water through the thermopile flow sensor (Fig. 7). The flow is restricted to a few µl·min$^{-1}$ by the hydraulic resistance of the surface microchannels. The elevation head gives a stable flow rate, which is determined accurately by weighing. The volume flow rate is calculated by taking the time-derivative of the amount of liquid flowing onto a scale, divided by the density of water.

The power feedback loop and PI controller were implemented in MATLAB. An Hewlett Packard 34401A multimeter was used to measure the zero-offset thermopile output voltage $V_{TC}$. The heating powers $P_1$, $P_2$ dissipated in the resistors are controlled by a Keithley 2602 two-channel source measurement unit (SMU). The SMUs control the dissipated power in the heater resistors by measuring the voltage and adjusting the current through the resistors. In this way the exact resistance value is not important and may drift.

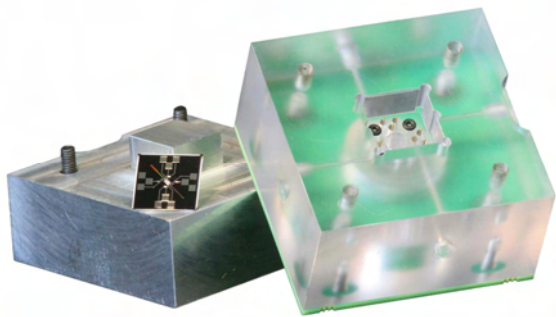

Fig. 6. Chip holder for fluidic and electrical interfacing with the sensor chip, measuring $12.5 \times 12.5$ mm.

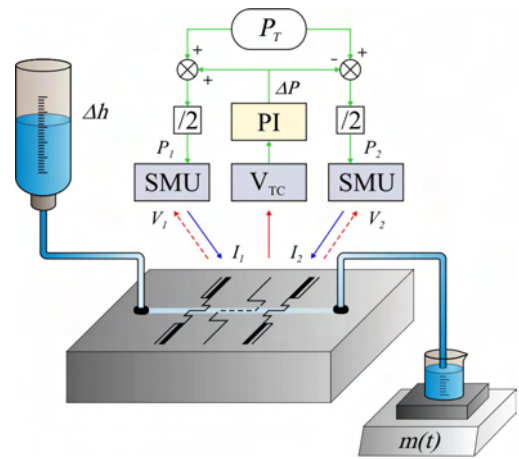

Fig. 7. Schematic overview of experimental setup, applying temperature-balancing power feedback control with $\Delta P/P_T$ being the flow dependent sensor output.

The total heating power $P_T$ was kept constant at 0.1 mW during these experiments, resulting in a temperature rise of a few Kelvin of the heater resistors with respect to the substrate. The difference in heating power $\Delta P$ needed to cancel the thermopile output voltage $V_{TC}$ was measured for applied water volume flow rates $Q$ up to 1.5 µl·min$^{-1}$. Fig. 8 shows the measured $\Delta P/P_T$ ratio together with results obtained from FVM model calculations. FVM model and measurement results show the sensor having a linear response, with 0.89 µl·min$^{-1}$ sensitivity $S_{\Delta P/P_T}$, up to about 0.5 µl·min$^{-1}$.

Fig. 9 shows the control system response on applying a stepwise changing volume flow rate. It shows the measured heating powers $P_1$, $P_2$ being adjusted in response to the flow rate. Correct operation of controlling the dissipated powers in the heater resistors is demonstrated by the fact that a change in the resistance values $R_1$ and $R_2$, mostly due to a change in $\Delta P$, is not visible in the output signal. The output signal $\Delta P/P_T$ is a measure of the flow rate, with the thermopile output voltage $V_{TC}$ being controlled back to zero after every step in the flow rate.

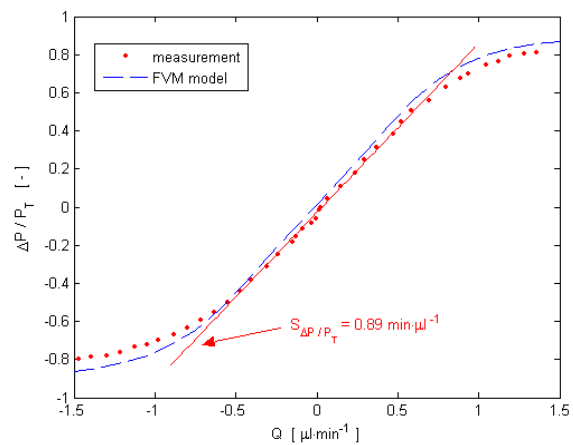

Fig. 8. Measured sensor output $\Delta P/P_T$ as a function of water volume flow rate $Q$. The total heating power $P_T$ is kept constant at 0.1 mW.

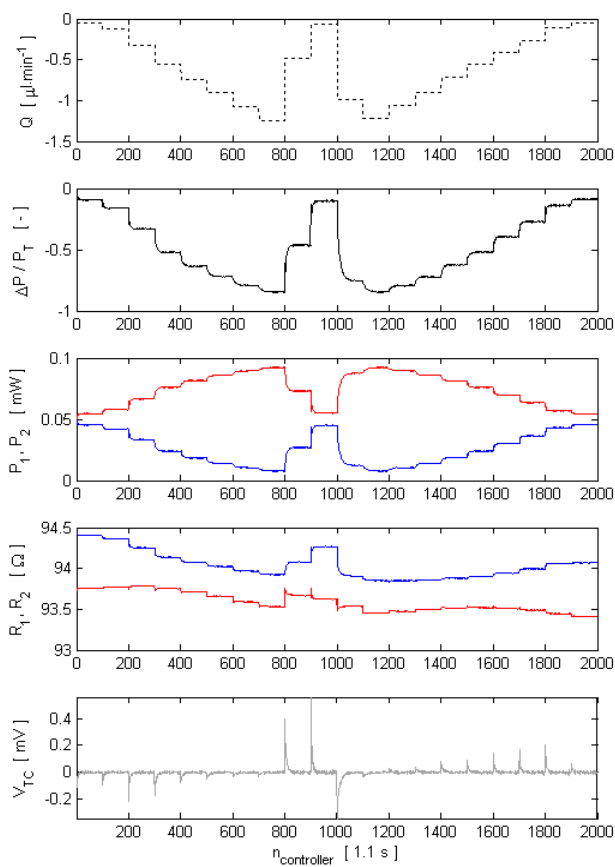

Fig. 9. Measured sensor control system response at changing flow rate. The total heating power $P_T$ is kept constant at 0.1 mW.

The low-drift operation of the temperature-balancing control system was verified by controlling $\Delta P$ for 13 hrs with no flow applied, while dissipating 1 mW total power $P_T$. A Hewlett Packard 34420A nanovolt meter was used to measure the thermopile output voltage $V_{TC}$. Fig. 10 shows obtained normalised power spectral densities. The power spectral densities of the average of both heater resistors $R_\mu$ and the difference between both heater resistors $R_\Delta$ show 1/f noise. The drift in $R_\mu$ can be mostly attributed to fluctuations in ambient temperature, while drift in $R_\Delta$ is, for a large part, caused by resistance drift. The integral action of the PI controller cancels low frequency drift in the thermopile output voltage $V_{TC}$, resulting in a smaller power spectral density at lower frequencies. The power spectral density for $\Delta P$ nearly resembles white noise, moreover $\Delta P$ and $V_{TC}$ are Gaussian distributed, demonstrating the correct low-drift operation of the temperature-balancing control system, eliminating influences of resistance drift. However, the distribution for $\Delta P$ shows a small offset, which can only be caused by heat-conduction asymmetries in the sensor or internal offsets in the measurement equipment. Neither of which should exhibit any significant drift over long time periods.

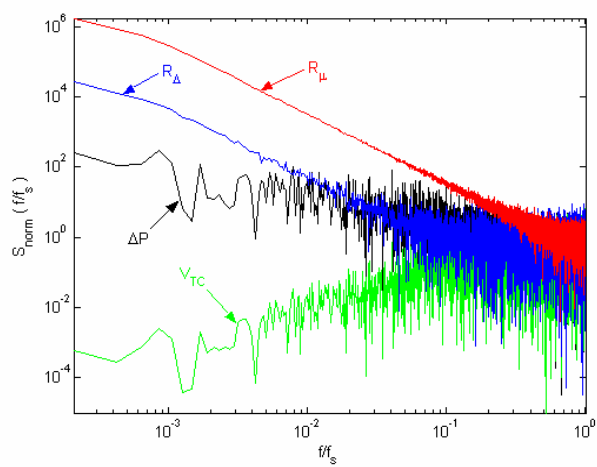

Fig. 10. Drift measurement for 13 hrs with 0.24 Hz sampling frequency $f_S$. Normalised power spectral density $S_{norm}$ is shown for the average of both heater resistors $R_\mu$, the difference between both heater resistors $R_\Delta$, the thermopile voltage $V_{TC}$ and the required power difference $\Delta P$, while dissipating 1 mW total power $P_T$.

## V. CONCLUSIONS

A thermally-isolated microchannel flow sensor has been fabricated, with integrated Al/poly-Si$^{++}$ thermopile and up- and downstream Al heater resistors. Flow measurements and FVM model calculations have shown that a temperature-balancing control system, which cancels the flow-induced temperature difference across the thermopile by controlling a power difference between both heater resistors, can give a linear measure of water flow rates up to 0.5 µl·min$^{-1}$. More importantly, the inherently zero-offset drift-free voltage output of the thermopile has been exploited to eliminate influences in resistance drift and mismatch of thin-film metal resistors used.

### ACKNOWLEDGEMENT

The authors would like to thank the Dutch Technology Foundation (STW) for financial support through the low-drift micro-flowsensors project (TET.6634).

### REFERENCES

[1] M. Dijkstra, M.J. de Boer, J.W. Berenschot, T.S.J. Lammerink, R.J. Wiegerink, M. Elwenspoek, *"Miniaturized Flow Sensor with Planar Integrated Sensor Structures on Semicircular Surface Channels"*, Proceedings of IEEE MEMS, 2007

[2] Y. Mizuno, M. Liger, Y-C. Tai, *"Nanofluidic Flowmeter Using Carbon Sensing Element"*, Proceedings of IEEE MEMS, 2004

[3] T.S.J. Lammerink, N.R. Tas, G.J.M. Krijnen, M. Elwenspoek, *"A New Class of Thermal Flow Sensors Using ΔT=0 as a Control Signal"*, Proceedings of IEEE MEMS, 2000

[4] P. Bruschi, A. Diligente, D. Navarrini, M. Piotto, *"A Double Heater Integrated Gas Flow Sensor with Thermal Feedback"*, Sensors and Actuators A., Vol. 123-124, 2005

[5] M. Dijkstra, M.J. de Boer, J.W. Berenschot, T.S.J. Lammerink, R.J. Wiegerink, M. Elwenspoek, *"A Versatile Surface Channel Concept for Microfluidic Applications"*, Journal of Micromechanics and Microengineering, Vol. 17, No. 10, 2007